\documentclass[journal]{IEEEtran}
\usepackage[T1]{fontenc}
\usepackage{cite}
\usepackage{url}
\usepackage{amsmath,amssymb,amsfonts}
\usepackage{graphicx}
\usepackage{textcomp}
\usepackage[table]{xcolor}
\usepackage{bm}
\usepackage{amsthm}
\usepackage{enumerate}
\usepackage{booktabs}
\usepackage{diagbox}
\usepackage{supertabular}
\usepackage{subfigure}
\usepackage{caption}
\usepackage{algorithmicx}
\usepackage[linesnumbered,ruled,vlined]{algorithm2e}
\usepackage{comment}  
\usepackage{filecontents}
\usepackage{flushend}
\usepackage{multirow}
\usepackage{array}
\definecolor{seagreen}{rgb}{0.18, 0.55, 0.34}
\definecolor{royalpurple}{rgb}{0.47,0.32,0.66}
\definecolor{brown(traditional)}{rgb}{0.59, 0.29, 0.0}
\definecolor{blue}{rgb}{0.3, 0.2, 0.9}
\usepackage[colorlinks,
            linkcolor=blue,
            anchorcolor=blue,
            citecolor=blue]{hyperref}

\DeclareMathOperator*{\argmin}{argmin}

\begin{document}

\title{ParaVul: A Parallel Large Language Model and Retrieval-Augmented Framework for Smart Contract Vulnerability Detection}

\author{
Tenghui Huang, Jinbo Wen, Jiawen Kang, \textit{Senior Member, IEEE}, Siyong Chen, Zhengtao Li, Tao Zhang,\\ Dongning Liu, \textit{Senior Member, IEEE}, Jiacheng Wang, Chengjun Cai, Yinqiu Liu, and Dusit Niyato, \textit{Fellow, IEEE}
\thanks{

T. Huang, J. Kang, S. Chen, and Z. Li are with the School of Automation, Guangdong University of Technology, Guangzhou 510006, China (e-mails: 3123000938@mail2.gdut.edu.cn, kavinkang@gdut.edu.cn, 3122000875@mail2.gdut.edu.cn, 3123000988@mail2.gdut.edu.cn).

J. Wen is with the College of Computer Science and Technology, Nanjing University of Aeronautics and Astronautics, Nanjing 210016, China (e-mail: jinbo1608@nuaa.edu.cn).

T. Zhang is with the School of Cyberspace Science and Technology, Beijing Jiaotong University, Beijing 100044, China (e-mail: taozh@bjtu.edu.cn).

D. Liu is with the School of Computer Science and Technology, Guangdong University of Technology, Guangzhou 510006, China (e-mail: liudn@gdut.edu.cn).

J. Wang, Y. Liu, and D. Niyato are with the College of Computing and Data Science, Nanyang Technological University, Singapore (e-mails: jiacheng.wang@ntu.edu.sg, yinqiu001@ntu.edu.sg, dniyato@ntu.edu.sg).

C. Cai is with the Department of Computer Science, City University of Hong Kong, Hong Kong (e-mail: chengjun.cai@cityu-dg.edu.cn).
}
\thanks{\textit{Corresponding author: Jiawen Kang.}}
}
\maketitle

\begin{abstract}
Smart contracts play a significant role in automating blockchain services. Nevertheless, vulnerabilities in smart contracts pose serious threats to blockchain security. Currently, traditional detection methods primarily rely on static analysis and formal verification, which can result in high false-positive rates and poor scalability. Large Language Models (LLMs) have recently made significant progress in smart contract vulnerability detection. However, they still face challenges such as high inference costs and substantial computational overhead. In this paper, we propose ParaVul, a parallel LLM and retrieval-augmented framework to improve the reliability and accuracy of smart contract vulnerability detection. Specifically, we first develop Sparse Low-Rank Adaptation (SLoRA) for LLM fine-tuning. SLoRA introduces sparsification by incorporating a sparse matrix into quantized LoRA-based LLMs, thereby reducing computational overhead and resource requirements while enhancing their ability to understand vulnerability-related issues. We then construct a vulnerability contract dataset and develop a hybrid Retrieval-Augmented Generation (RAG) system that integrates dense retrieval with Best Matching 25 (BM25), assisting in verifying the results generated by the LLM. Furthermore, we propose a meta-learning model to fuse the outputs of the RAG system and the LLM, thereby generating the final detection results. After completing vulnerability detection, we design chain-of-thought prompts to guide LLMs to generate comprehensive vulnerability detection reports. Simulation results demonstrate the superiority of ParaVul, especially in terms of F1 scores, achieving $0.9398$ for single-label detection and $0.9330$ for multi-label detection.
\end{abstract}


\section{Introduction}

\IEEEPARstart{W}{\MakeLowercase{ith}} the advancement of distributed ledger technologies such as blockchain~\cite{zheng2018blockchain, 10254627} and Web 3.0~\cite{10816471, 10542397}, smart contracts have become an indispensable foundation for decentralized finance ecosystems~\cite{liu2025sok}. They can automatically execute and verify transactions based on predefined conditions, significantly enhancing transaction efficiency. However, once deployed on blockchains, the immutable nature of smart contracts can expose potential vulnerabilities to malicious exploitation, leading to significant security risks and economic losses~\cite{11037432}. Notably, high-profile security incidents, such as the decentralized autonomous organization attack~\cite{mehar2019understanding}, underscore the severe consequences of these vulnerabilities, which have attracted attention from academia and industry~\cite{10.1145/3769013}. Therefore, smart contract vulnerability detection is crucial for the security of the blockchain ecosystem~\cite{10.1145/3750042}.

Traditional methods for smart contract vulnerability detection typically depend on static analysis, dynamic detection, and formal verification~\cite{10.1145/3750042, 10.1145/3769013}. Although these methods are capable of detecting certain contract security issues, they still face several critical limitations, including low efficiency, inability to meet real-time analysis demands for large-scale contracts, limited generalization and adaptability to emerging vulnerability types, and insufficient understanding of complex contract logic~\cite{10.1145/3769013, feist2019slither}. These limitations can result in high false-positive rates~\cite{10.1145/3750042}. As smart contract applications become more complicated, the limitations of traditional methods in addressing new vulnerabilities become more obvious~\cite{10.1145/3750042}.

Recent advances in Large Language Models (LLMs)~\cite{chen2025chatgpt,liu2024propertygpt} and Retrieval-Augmented Generation (RAG) technologies in natural language processing offer new approaches for smart contract vulnerability detection~\cite{liu2024propertygpt,boi2024smart,yu2024retrieval}. On the one hand, LLMs are capable of directly understanding and analyzing smart contracts to identify semantic security vulnerabilities. The prominent advantages of LLMs in smart contract vulnerability detection lie in their ability to understand complex contract logic, adapt to new types of vulnerabilities, and generate explanations and repair suggestions, significantly enhancing both the intelligence and practicality of detection. On the other hand, RAG retrieves similar cases from the vulnerability knowledge base to enhance the identification capability of LLMs. The applications of LLMs in smart contract vulnerability detection have attracted increasing attention. For instance, the authors in~\cite{sun2024gptscan} combined LLMs with control-flow graph analysis to enhance detection accuracy. Similarly, RAG technology has also been applied to smart contract vulnerability detection. For instance, the authors in~\cite{yu2024retrieval} integrated a vector database of vulnerable contracts with LLMs to enhance detection. However, existing studies still suffer from several limitations, including high computational cost, reliance on high-quality datasets, and considerable false-positive rates. Moreover, directly applying these technologies to smart contract vulnerability detection remains challenging due to insufficient code semantic understanding, high resource consumption, and poor generalization.

To this end, in this paper, we propose ParaVul, a parallel LLM and retrieval-augmented framework for smart contract vulnerability detection. In this framework, we first develop Sparse Low-Rank Adaptation (SLoRA) to reduce the computational overhead of LLM fine-tuning, thereby enhancing the performance of the LLM in vulnerability detection. We then propose a hybrid RAG system to verify the detection results generated by the LLM. Unlike single RAG systems, the hybrid RAG system filters results through multiple retrieval strategies. Furthermore, we feed the filtered results, together with the outputs of the LLM, into a meta-learning model~\cite{lin2021metagater}, which performs weighted processing to generate the final detection results. Finally, we utilize Chain-of-Thought (CoT) prompt techniques to guide the LLM to create detailed and accurate vulnerability reports. These reports provide an in-depth analysis of the detected vulnerabilities, assisting auditors in understanding their specific characteristics. The main contributions of this paper are summarized as follows:

\begin{itemize}
\item We propose ParaVul, a novel smart contract vulnerability detection framework that integrates an LLM with RAG to analyze smart contracts and identify potential vulnerabilities efficiently. ParaVul employs parallel processing to synchronize LLM-based detection with RAG-based detection, effectively enhancing detection accuracy. Moreover, we design a vulnerability detection report template to help users clearly understand identified vulnerabilities and corresponding remediation suggestions.

\item We develop SLoRA based on Quantized LoRA (QLoRA) to reduce the computational overhead of LLM fine-tuning while enhancing detection performance. Specifically, we dynamically remove non-critical connections and freeze the backbone parameters of the LLM, training only the adapter layers. Through this design, SLoRA is capable of improving LLM performance in smart contract vulnerability detection while reducing computational cost.

\item To mitigate hallucinations in LLM-based vulnerability detection, we construct a vulnerability contract dataset and develop a hybrid RAG system that integrates dense retrieval with Best Matching 25 (BM25). By retrieving relevant vulnerability samples from the database through these complementary strategies, the RAG system provides auxiliary validation for the detection results of the LLM, thereby improving detection reliability.

\item We propose a verification module that leverages a meta-learning model to refine the final results of smart contract vulnerability detection. By aggregating the outputs of LLM and RAG detection, the meta-learning model constructs a fusion feature vector and trains a Multi-Layer Perceptron (MLP) as the meta-learner, enabling accurate vulnerability identification. This verification module can adaptively integrate the advantages of LLMs and RAG, thereby improving overall detection performance. In terms of F1-scores, the verification module achieves at least a $4\%$ improvement over LLMs and a $12\%$ improvement over RAG.
\end{itemize}

The remainder of this paper is structured as follows: Section \ref{section: II} reviews the related work. In Section \ref{section: III}, we propose ParaVul. Sections \ref{section: IV}, \ref{section: V}, and \ref{section: VI} present the designs of SLoRA, the hybrid RAG system, and the verification module, respectively. Section \ref{section: VII} evaluates the performance of ParaVul. Section \ref{section: VIII} concludes the paper. Key mathematical notations of this paper are illustrated in Table \ref{Notation_Table}.

\begin{table}[t]
    \renewcommand{\arraystretch}{1.5}
    \caption{Key Mathematical Notations of this Paper}
    \label{Notation_Table}
    \resizebox{\columnwidth}{!}{
    \begin{tabular}{m{0.8cm}|m{7.0cm}} 
    \toprule[1.5pt]
     \rowcolor{gray!10}
    \hline
    \multicolumn{1}{c|}{\textbf{Notations}}  & \multicolumn{1}{c}{\textbf{Definition}} \\ \hline
        $B$ & Batch size \\\hline
        $\boldsymbol{C}$ & Contracts in dataset \\\hline
        $\mathcal{D}$ & Vulnerable smart contract dataset \\\hline
        $f_{\mathrm{ID}}$ & Inverse document frequency \\\hline
        $f_{{T}}$ & Term frequency \\\hline
        $k$ & Number of nonzero entries to retain in $\boldsymbol{S}$ \\\hline
        $l$ & Document length \\\hline
        $\bar{l}$ & Average document length \\\hline
        $\boldsymbol{L}$ & Vulnerability labels on smart contract \\\hline
        $\boldsymbol{M}$ & Binary mask matrix, $\boldsymbol{M}_{ij}\in\{0,1\}$ \\\hline
        $n$ & Number of terms in the query \\\hline
        $\boldsymbol{q}$ & Query of smart contract codes \\\hline
        $r$ & LoRA rank \\\hline
        $\boldsymbol{S}$ & Trainable sparse matrix \\\hline
        $T$ & Number of epochs \\\hline
        $\tau$ & Threshold value, the $k$-th largest entry of $|\boldsymbol{S}|$ \\\hline
        $\mathbf{\alpha}$ & Sparsity level \\\hline
        $\mathbf{\eta}$ & Learning rate of adapters \\\hline
    \bottomrule[1.5pt]
    \end{tabular}\label{Notation}
    }
\end{table}

\section{Related Work}
\label{section: II}


In this section, we review the related work across three domains: traditional smart contract vulnerability detection, LLM-based vulnerability detection, and sparse optimization of adapter fine-tuning. The comparison of the current literature and this paper is summarized in Table \ref{comparison}.

\begin{table*}[t]
\centering
\caption{Comparison Between the Current Literature and This Paper}
\label{comparison}
\renewcommand{\arraystretch}{1.3}
\setlength{\tabcolsep}{4pt}

\resizebox{\textwidth}{!}{
\begin{tabular}{c|c|cccccccccccc}
\toprule[1.5pt]
\hline
\rowcolor{gray!10}
\multicolumn{2}{c|}{\textbf{Literature}} 
& \cite{feist2019slither} & \cite{choi2021smartian} & \cite{abdellatif2018formal} & \cite{sun2024gptscan} & \cite{choi2025smart} & \cite{yu2024retrieval}  & \cite{kevin2025smartllm} & \cite{jin2025llm} & \cite{ding2023sparse} & \cite{hu2025lors} & \cite{han2024sltrain} & \textbf{Our Paper} \\
\midrule
\multirow{3}{*}{\textbf{\shortstack{Optimization\\Objectives}}} 
& Detection Performance 
& \checkmark & \checkmark & \checkmark & \checkmark & \checkmark 
& \checkmark & \checkmark & \checkmark 
&  &  &  & \textcolor{red}{\checkmark} \\
& Computational Resource    
&  &  &  &  &  
&  &  & 
& \checkmark & \checkmark & \checkmark & \textcolor{red}{\checkmark} \\
& Latency    
& &  & & & &  
& &  & & & & \textcolor{red}{\checkmark} \\
\midrule
\multirow{3}{*}{\textbf{Solutions}} 
& Traditional
& \checkmark & \checkmark & \checkmark &  & 
& & &  
&  &  &  & \textcolor{red}{} \\
& LLM-based    
& & &  &\checkmark & \checkmark
& \checkmark & \checkmark & \checkmark
&  \checkmark & \checkmark & \checkmark & \textcolor{red}{\checkmark} \\
& RAG-based    
& & &  & &  &  \checkmark
& \checkmark & \checkmark & & & & \textcolor{red}{\checkmark} \\
\hline
\bottomrule[1.5pt]
\end{tabular}
}
\end{table*}

\subsection{Traditional Smart Contract Vulnerability Detection}

Conventional approaches to smart contract vulnerability detection primarily consist of static analysis, dynamic analysis, and formal verification~\cite{10.1145/3769013}. Static analysis discovers potential vulnerabilities by checking the source code or bytecode of smart contracts. For example, the authors in~\cite{feist2019slither} proposed a static analysis framework designed to efficiently provide detailed information about Ethereum smart contracts. Dynamic analysis identifies vulnerabilities by executing smart contracts and observing their runtime behaviors. In~\cite{choi2021smartian}, the authors proposed a practical open-source fuzzer, which statically analyzes smart contract bytecodes to predict effective transaction sequences. The proposed fuzzer also identifies constraints that each transaction must satisfy. Dynamic analysis can also discover potential vulnerabilities related to runtime performance~\cite{kushwaha2022ethereum}. Meanwhile, formal verification leverages mathematical methods to rigorously prove the security of smart contracts~\cite{almakhour2023formal}. For instance, the authors in~\cite{abdellatif2018formal} proposed a formal modeling approach to verify smart contract behaviors within the execution environment. However, these traditional approaches face inherent limitations when addressing complex vulnerabilities and emerging attack vectors.


\subsection{LLM-based Vulnerability Detection}

With the development of LLMs, their applications in smart contract vulnerability detection have emerged as a research hotspot. For example, the authors in~\cite{sun2024gptscan} proposed a method that integrates LLMs with program analysis to detect logic vulnerabilities. In~\cite{chen2025chatgpt}, the authors conducted a systematic assessment of the capabilities and limitations of LLMs in vulnerability detection. The authors in~\cite{choi2025smart} proposed a method that combines LLMs with control-flow graph context fusion analysis, leveraging dual-modal features to improve both accuracy and localization precision. However, the applications of LLMs in vulnerability detection still face challenges, such as the need for high-quality datasets, the interpretability of models, and the consumption of computational resources.

The applications of RAG in the field of smart contract vulnerability detection are also an emerging topic. The authors in~\cite{yu2024retrieval} created a vector database involving 830 known vulnerability contracts and combined it with GPT to build a RAG system for smart contract vulnerability detection. In~\cite{kevin2025smartllm}, the authors proposed a RAG-enhanced LLM framework by utilizing QLoRA to fine-tune LLMs. Moreover, the authors combined the proposed framework with a standard library knowledge base to perform contextual inference and vulnerability detection. The authors in~\cite{jin2025llm} proposed a method that merges a three-stage decompose-retrieve-generate pipeline with multi-agent collaboration. However, these works still suffer from high false-positive rates.


\subsection{Sparse Optimization of Adapter Fine Tuning}

To mitigate the high computational demands of LLMs, sparsification techniques have been explored on top of LoRA~\cite{ding2023sparse, hu2025lors,han2024sltrain}. In~\cite{ding2023sparse}, the authors proposed sparse LoRA, which combines LoRA with a sparse gating mechanism. In~\cite{hu2025lors}, the authors proposed a LoRA-based method for sparse LLMs, which combines sparse structure preservation with LoRA injection to achieve effective sparsification. Similarly, in~\cite{han2024sltrain}, the authors proposed a low-rank unified modeling method that balances compression and acceleration through trainable sparse structures, thereby enabling efficient sparsification of LLMs. Although these adapter-based sparsification approaches achieve computational compression in large-scale model training environments, their complexity results in extremely low transferability. To this end, we propose SLoRA, which integrates a sparse matrix into QLoRA-based LLMs to reduce computational overhead, while ensuring both architectural simplicity and high transferability.


\section{Framework Design}
\label{section: III}

\begin{figure*}
    \centering
    \includegraphics[width=0.98\textwidth]{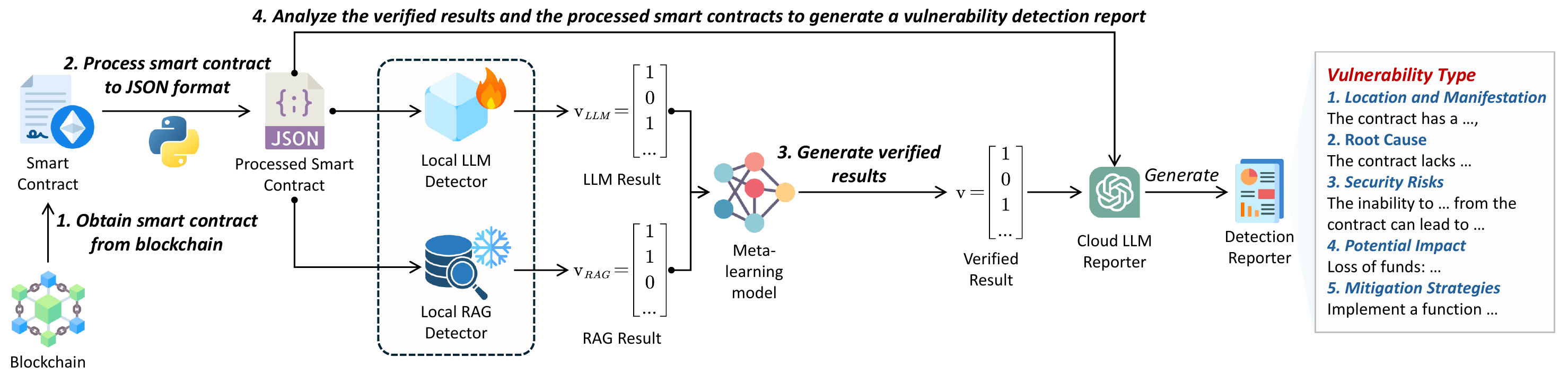}
    \caption{The architecture of ParaVul, which is an intelligent framework for smart contract vulnerability detection that leverages parallel LLMs and RAG. In ParaVul, smart contracts are directly sourced from the blockchain. To ensure detection efficiency, we locally deploy an LLM, the hybrid RAG system, as well as the verification module. Finally, we use API calls to access a cloud-based LLM to generate comprehensive vulnerability detection reports.}
    \label{Framework}
\end{figure*}

In this section, we introduce ParaVul, a framework consisting of four stages: data preprocessing, parallel detection using LLMs and the hybrid RAG system, detection result verification, and report generation. These stages are tightly integrated to ensure both the efficiency of the detection process and the accuracy of the detection results. The architecture of ParaVul is shown in Fig. \ref{Framework}.


\subsection{Data Preprocessing}
At this stage, we preprocess the original smart contract code to ensure that the input data used for subsequent detection is both high-quality and semantically complete. Specifically, we standardize the smart contract code and convert it into a unified JSON format. Each smart contract is then represented by a binary vector~\cite{huang2022smart}, where each element ($0$ or $1$) indicates the absence or presence of a specific vulnerability type. This binary vector serves as the label representation for the corresponding contract in our dataset. At the same time, we perform noise elimination, remove redundant information and non-critical comments, and retain the complete semantic context, thereby ensuring that LLMs and the hybrid RAG system can comprehensively capture code security information.

\subsection{LLM and RAG Parallel Detection}
LLMs leverage robust code semantic understanding and contextual reasoning capabilities to automatically identify potential logical flaws and unsafe implementations~\cite{wang2024sanitizing}. Meanwhile, RAG integrates external security knowledge bases and specifications, ensuring that detection results are both accurate and traceable~\cite{yu2023retrieval}. In the following, we present ParaVul, which leverages these two detection paths in parallel to fully exploit their respective advantages and achieve comprehensive identification of smart contract vulnerabilities.

\begin{itemize}
\item \textbf{LLM Detector:} We locally deploy an LLM and fine-tune it using SLoRA. The model takes structured smart contracts as input and outputs vector representations corresponding to the detected vulnerability types.

\item \textbf{RAG Detector:} We locally deploy the proposed hybrid RAG system, which consists of two components: a dense retrieval and a BM25 retrieval. These two retrieval strategies employ distinct data preprocessing pipelines and voting mechanisms.

\end{itemize}

The processed smart contract code is simultaneously analyzed by both LLM and RAG detectors. Their detection outputs are subsequently vectorized~\cite{devlin2019bert}, enabling structured comparison and comprehensive quantitative analysis~\cite{kornblith2019similarity}. The vector representations ensure consistency across the two detectors and facilitate efficient integration of results into the overall framework~\cite{liu2022deep}, while parallel detection significantly reduces the overall processing time.

\subsection{Detection Result Verification}

To ensure the reliability of detection results, we design a verification module based on a meta-learning model, which refines the final results of smart contract vulnerability detection. This module uses the detection results from both the LLM and RAG detectors as input. By leveraging a meta-learning model~\cite{lin2021metagater}, it can rapidly adapt to new tasks while learning the importance weights of different features. Through this verification module, the outputs of the two detectors are weighted and aggregated to produce the final detection results. By jointly considering the strengths of both LLM and RAG detectors, the verified results achieve higher accuracy and robustness in vulnerability detection.

\subsection{Report Generation}

The verified detection results then enter the report generation stage, which aims to provide users with intuitive, comprehensive, and practically instructive vulnerability detection reports. To this end, we design a CoT-based report template tailored for smart contract vulnerability detection. Unlike ordinary reports that merely list vulnerability types~\cite{sendner2023smarter,10.1145/3560905.3568175}, our detection report offers a structured overview of the detected vulnerabilities, comprising the following five elements:

\subsubsection{Location and manifestation}
A detailed description of the specific location and manifestation of vulnerabilities within the smart contract. For example, a fallback function may permit token deposits but lack a withdrawal mechanism~\cite{jiao2024survey}. This vulnerability can cause tokens to become permanently locked within the smart contract.

\subsubsection{Root causes}
The report provides an in-depth analysis of the root causes of vulnerabilities. For instance, it may highlight that the smart contract lacks functionality for withdrawing or transferring Ether, which is the native cryptocurrency of the Ethereum blockchain~\cite{tsankov2018securify}. Furthermore, it may indicate scenarios where Ether becomes irretrievable from the smart contract under certain conditions.

\subsubsection{Security risks}
The report explicitly outlines the potential security risks associated with vulnerabilities. For instance, it can emphasize the risk of permanent fund loss if Ether is inadvertently sent to the smart contract address or if the transfer operation of the owner fails.

\subsubsection{Potential impact}
The report specifies the potential impact of the vulnerability, noting that blockchain tokens sent to the smart contract address may become unrecoverable, potentially resulting in financial loss.

\subsubsection{Mitigation strategies}
The report outlines concrete mitigation strategies. For instance, it can recommend the implementation of a withdrawal function that allows the smart contract owner to retrieve blockchain tokens, thereby ensuring that any tokens sent to the contract remain recoverable.

\begin{figure}
    \centering
    \includegraphics[width=0.98\linewidth]{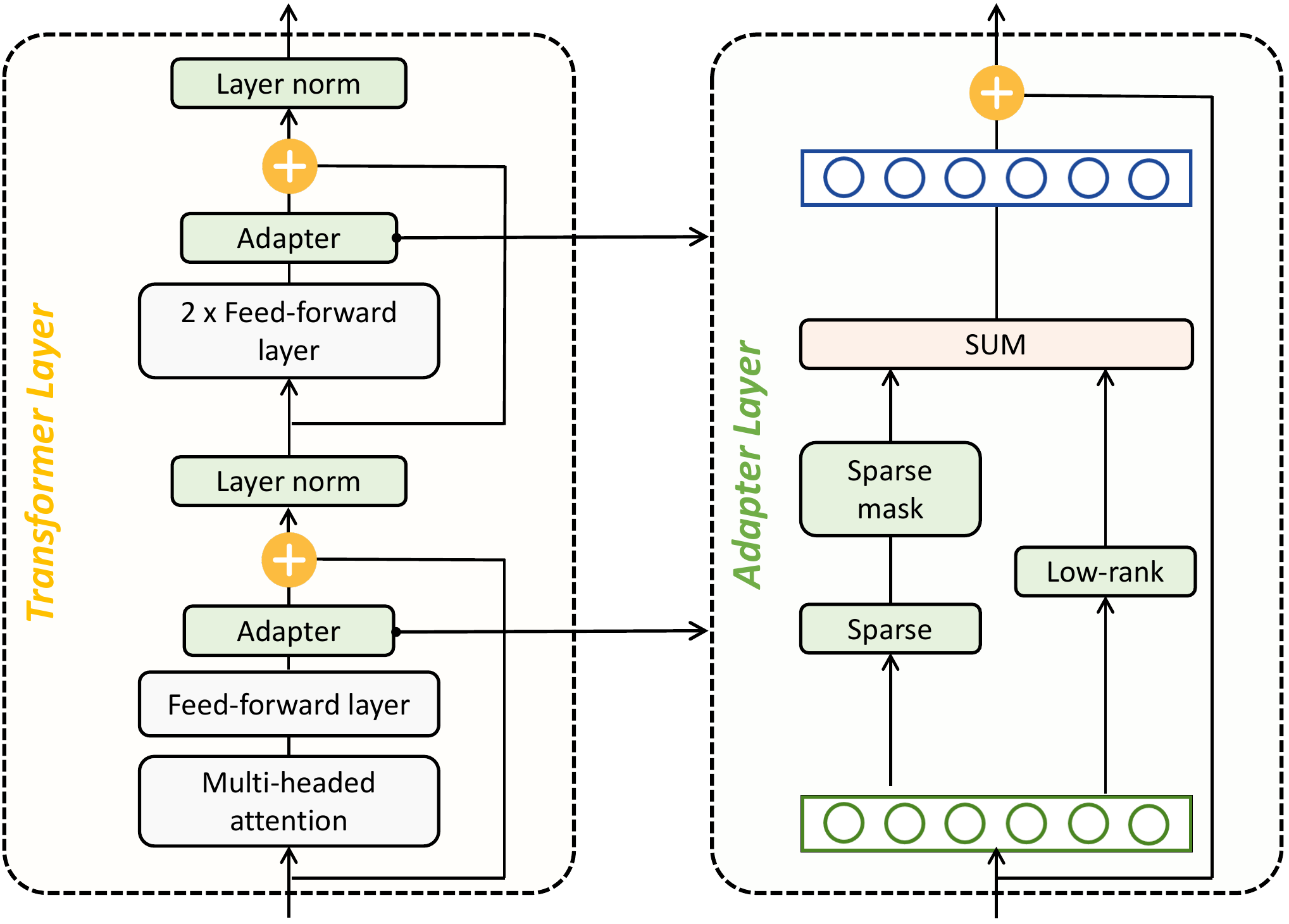}
    \caption{The architecture of SLoRA, which incorporates two adapter modules into each transformer layer: one placed after the projection layer following multi-head attention, and the other after the two fully connected layers. Each adapter consists of a sparse layer and a LoRA layer. The sparse layer masks a subset of nodes, while the LoRA layer applies a low-rank matrix transformation to the data.}
    \label{SLoRA}
\end{figure}

\begin{algorithm}[t]
\label{adapter_algorithm}
\DontPrintSemicolon
\SetAlgoLined

Initialize frozen quantized weights $W_q$.  

Initialize low-rank parameters $\boldsymbol{U} \in \mathbb{R}^{d \times r}, \boldsymbol{V} \in \mathbb{R}^{r \times d}$.  

Initialize sparse matrix $\boldsymbol{S} \in \mathbb{R}^{d \times d}$.  

Initialize binary mask $\boldsymbol{M} \in \mathbb{R}^{d \times d}$.  

\For{\rm{each training iteration}}
{
    \textcolor{blue}{\textit{\#\#\# Low-Rank Adapter \#\#\#}}
    
    Compute the low-rank increment: $\Delta W_{\mathrm{low}} = \boldsymbol{U}\boldsymbol{V}$.  
    
    Obtain the low-rank output: $O_{\mathrm{low}} = \boldsymbol{x}\,(\boldsymbol{U}\boldsymbol{V})$.  

    \textcolor{blue}{\textit{\#\#\# Sparse Adapter \#\#\#}}
    
    Calculate active entries: $k = \lfloor (1-\alpha)d^2 \rfloor$.  
    
    Select the top-$k$ elements of $\boldsymbol{S}$.  
    
    Construct the binary mask $\boldsymbol{M}$.  
    
    Obtain the sparse output: $O_{\mathrm{spr}} = \boldsymbol{x}\,(\boldsymbol{S} \odot \boldsymbol{M})$.  

    \textcolor{blue}{\textit{\#\#\# Combined Outputs \#\#\#}}
    
    Compute the base output: $O_{\mathrm{base}} = \boldsymbol{x}\,W_q$.  
    
    Compute the final output: $O = O_{\mathrm{base}} + O_{\mathrm{low}} + O_{\mathrm{spr}}$.  

    \textcolor{blue}{\textit{\#\#\# Frozen Base Parameters \#\#\#}}
    
    Keep base parameters fixed: $\tfrac{\partial \mathcal{L}}{\partial W_q} = 0$.  
    
    Update only $\boldsymbol{U}, \boldsymbol{V},\boldsymbol{S}$.  

    \textcolor{blue}{\textit{\#\#\# Loss Function \#\#\#}}
    
    Optimize with multi-label BCE loss:  
    
    $\mathcal{L} = -\tfrac{1}{L}\sum_{i=1}^L\big[y_i\log \hat{y}_i + (1-y_i)\log(1-\hat{y}_i)\big]$.  
}

\textbf{return} the adapter parameters $\boldsymbol{U}, \boldsymbol{V}, \boldsymbol{S}$.  

\caption{SLoRA for Smart Contract Vulnerability Detection by LLMs}\label{alg:1}
\end{algorithm}

\begin{figure*}
    \centering
    \includegraphics[width=0.98\textwidth]{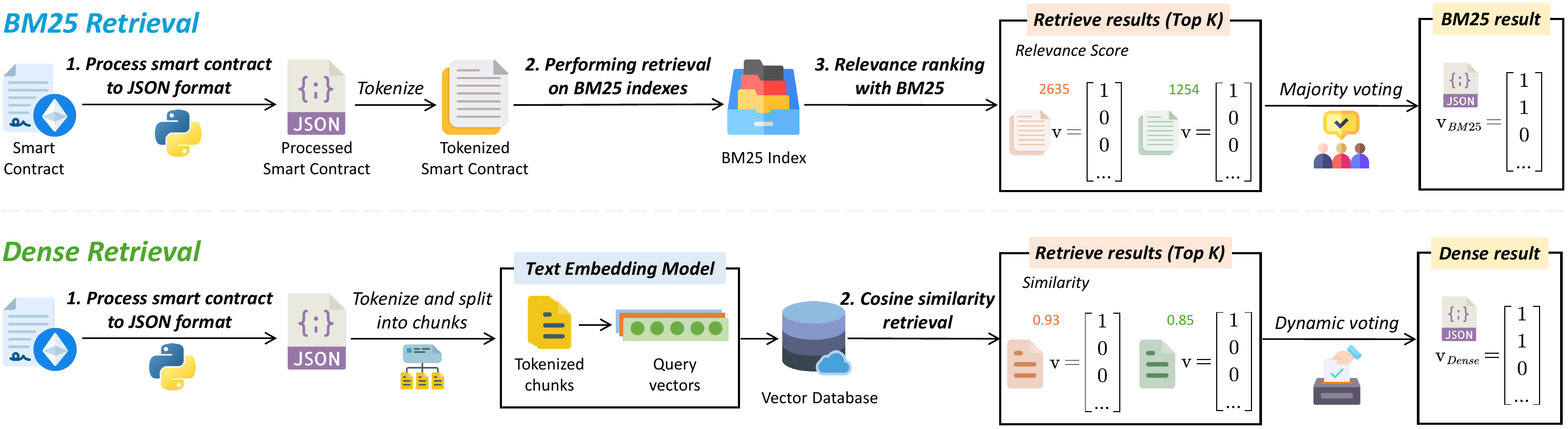}
    \caption{The architecture of the hybrid RAG system, which presents the vulnerability detection process corresponding to the BM25 retrieval strategy and the dense retrieval strategy, respectively.}
    \label{RAG_System}
\end{figure*}

\section{Sparse Low-Rank Adaptation}
\label{section: IV}
In this section, we introduce the developed SLoRA. The inputs to SLoRA include a pre-trained LLM, a smart contract dataset, and a set of hyperparameters, including batch size $B$, learning rate $\eta$, number of training epochs $T$, low-rank dimension, early stopping patience, and the sparsity ratio~\cite{dettmers2023spqr}. The objective of SLoRA is to produce a fine-tuned and sparsified model. Similar to LoRA, SLoRA freezes all layers except the adapter layer, and during fine-tuning, only the parameters of the adapter layer are updated, significantly reducing the computational requirements for LLM fine-tuning.

As shown in Fig.~\ref{SLoRA}, SLoRA enhances the quantized base model with two additional modules: a low-rank adapter and a sparse adapter. This design improves the expressive power of LLMs while minimizing the number of trainable parameters, thereby optimizing both performance and efficiency.

\subsection{Low-Rank Adapter}
    
Given an input feature matrix \(\boldsymbol{x} \in \mathbb{R}^{n\times d}\), where $n$ is the number of samples and $d$ is the feature dimension of each sample~\cite{hu2022lora}, we introduce a rank‑\(r\) decomposition of the weight increment as
\begin{equation}
    \Delta W_{\mathrm{low}} = \boldsymbol{U} \boldsymbol{V},
\end{equation}
where $\boldsymbol{U} \in \mathbb{R}^{d \times r}$ and $\boldsymbol{V} \in \mathbb{R}^{r \times d}$. The corresponding low‑rank output increment is given by
\begin{equation}\label{LoRA_output}
   O_{\mathrm{low}} = \boldsymbol{x}\,(\boldsymbol{U} \boldsymbol{V}).
\end{equation}

This decomposition leverages the low-rank structure to capture the essential features of the data in a lower-dimensional space, thereby reducing the number of parameters required for training and enhancing computational efficiency~\cite{hu2022lora}.

\subsection{Sparse Adapter}
    
We first define a trainable sparse matrix \(\boldsymbol{S} \in \mathbb{R}^{d \times d}\) with a sparsity level \(\mathbf{\alpha} \in [0,1]\). The number of entries to retain is determined by
\begin{equation}
        k = \Bigl\lfloor (1 - \mathbf{\alpha})\,d^2 \Bigr\rfloor.
\end{equation}
    
We then construct the binary mask $\boldsymbol{M} \in \mathbb{R}^{d \times d}$ as
\begin{equation}
    \boldsymbol{M}_{ij} = 
    \begin{cases}
            1 & \text{if } |\boldsymbol{S}_{ij}| \ge \tau, \\
            0 & \text{otherwise},
    \end{cases}
\end{equation}
where \(\mathbf{\tau}\) represents the \(k\)-th largest element of \(\lvert \boldsymbol{S}\rvert\). This mask retains only the top-$k$ elements of $\boldsymbol{S}$ in terms of the absolute value, enforcing sparsity in the adapter layer while ignoring less significant connections. Hence, similar to~(\ref{LoRA_output}), the sparse output increment is given by
\begin{equation}\label{sparse_output}
         O_{\mathrm{spr}} = \boldsymbol{x}\,( \boldsymbol{S} \odot  \boldsymbol{M}),
\end{equation}
where \(\odot\) denotes the Hadamard product between matrices.

This sparse adapter introduces sparsity into the transformer layer, reducing the number of trainable parameters and enhancing its capacity to focus on the most relevant features.

\subsection{Combined Output}
The final output of the adapter layer is obtained by summing the quantized base output with the incremental outputs from the low-rank and sparse adapters, as expressed by
\begin{equation}
    O = O_{\mathrm{base}} + O_{\mathrm{low}} + O_{\mathrm{spr}},
\end{equation}
where $O_{\mathrm{base}} = \boldsymbol{x} W_q$ denotes the output of the quantized base model. This combined formulation allows the transformer layer to capture both low-rank and sparse structures, improving representation capacity while keeping the number of trainable parameters minimal.

\subsection{Frozen Base Model Parameters}

Considering the need to handle multiple labels simultaneously, we use the multi-label Binary Cross-Entropy (BCE) loss and define the training objective of SLoRA as
\begin{equation}\label{loss_function_SLoRA}
        \mathcal{L}
         = -\frac{1}{L}\sum_{i=1}^L
        \bigl[y_i\log \hat{y}_i + (1-y_i)\log(1-\hat{y}_i)\bigr],
\end{equation}
where \(L\) is the number of labels, $y_i \in \{0,1\}$ represents the binary label $i$, and \(\hat{y}_i\) is the predicted probability for label \(i\). Based on (\ref{loss_function_SLoRA}), we freeze the weight matrix \(W_q\) of the quantized transformer model during fine-tuning, which is given by
\begin{equation}
        \frac{\partial \mathcal{L}}{\partial W_q} = 0.
\end{equation}
Note that parameter updates are confined to \(\boldsymbol{U}\), \(\boldsymbol{V}\), and \(\boldsymbol{S}\). This strategy preserves the knowledge embedded in the quantized transformer model while allowing the adapters to learn task-specific refinements that enhance overall performance.


\subsection{Computational Complexity Analysis}

For the low-rank adapter, the computational complexity is $\mathcal{O}(n d r) + \mathcal{O}(n r d) = \mathcal{O}(n d r)$, which is significantly lower than the cost of a full-rank update $\mathcal{O}(n d^2)$~\cite{hu2022lora}. For the sparse adapter, the computational complexity is $O(n k)$, which is significantly lower than the dense case, particularly when the sparsity level $\alpha$ is close to $1$. Therefore, the overall computational complexity of SLoRA is expressed as
\begin{equation}
    \mathcal{O}(n (d r + k)) \ll \mathcal{O}(n d^2).
\end{equation}

In conclusion, SLoRA provides a principled approach for enhancing model expressivity while maintaining a minimal parameter footprint. By jointly leveraging low-rank and sparse adapters, SLoRA achieves a balanced trade-off between performance and computational efficiency, making it particularly suitable for resource-constrained applications. The procedural implementation of SLoRA is outlined in Algorithm \ref{alg:1}.

\section{Hybrid RAG System with Dense and BM25 Retrieval Strategies}\label{section: V}
In this section, we present the proposed hybrid RAG system, which integrates BM25 with dense retrieval to identify vulnerabilities in smart contracts. The architecture of the hybrid RAG system is illustrated in Fig. \ref{RAG_System}.

\subsection{BM25 Retrieval}

\subsubsection{Knowledge base construction}

The source code of each contract undergoes systematic preprocessing, including code cleaning, extraction of key components, tokenization with lowercase normalization, identification of security-related keywords, pattern-based feature extraction, and intelligent de-duplication. This process yields an optimized token list for each contract. We then construct a BM25 model based on the tokenized text, with term statistics initialized and computed~\cite{robertson2025bm25}. To ensure accurate mapping between BM25 indices and the original contracts, the order of preprocessed documents is strictly aligned with the training metadata.

\subsubsection{BM25 retrieval strategy}

Each query contract in the test set is first processed through word segmentation, and the resulting tokens are then fed into the constructed BM25 model. The BM25 model computes the relevance scores between the query and all smart contracts in the knowledge base, using the following formulation~\cite{robertson2025bm25}:
\begin{equation}
    \mathrm{BM25}(\boldsymbol{q},\boldsymbol{d})={\frac{\sum_{i=1}^{n}f_\mathrm{ID}(\boldsymbol{q}_i)\left(k_1+1 \right) f_{T}\left(\boldsymbol{q}_i,\boldsymbol{d}\right)}{k_1\left(1-b+\frac{l_D b}{\bar{l}}\right)+f_{T}\left(\boldsymbol{q}_i,\boldsymbol{d}\right)}},
\end{equation}
where $n$ denotes the number of terms in the query $\boldsymbol{q}$, $f_{T}(\boldsymbol{q}_i,\boldsymbol{d})$ represents the frequency of the term $\boldsymbol{q}_i$ in the document $\boldsymbol{d}$, $k_1$ is a free parameter that adjusts the term frequency scaling, and $b$ is another free parameter that controls the strength of length normalization. Moreover, $l_D$ denotes the length of the document $\boldsymbol{d}$, and $\bar{l}$ is the average document length in the collection. Finally, $f_\mathrm{ID}(\boldsymbol{q}_i)$ represents the inverse document frequency of the term $\boldsymbol{q}_i$, which is calculated as
\begin{equation}
    f_\mathrm{ID}(\boldsymbol{q}_i)=\log \left( \frac{N-n_{\boldsymbol{q}_i}+0.5}{\,n_{\boldsymbol{q}_i}+0.5}+1 \right),
\end{equation}
where $N$ represents the total number of contracts in the dataset, and $n_{\boldsymbol{q}_i}$ represents the number of contracts that contain the term $\boldsymbol{q}_i$. After computing the relevance scores, the BM25 model ranks all contracts in descending order and selects the most relevant ones. To ensure the objectivity and validity of the retrieval results, any document corresponding to the query contract itself is excluded during this process.

\subsection{Dense Retrieval}

\subsubsection{Knowledge base construction}
Since the source code of smart contracts often exceeds the maximum input sequence length, we adopt an advanced sliding window segmentation mechanism~\cite{hirzel2017sliding} to address this issue. This mechanism is designed to mitigate the loss of critical information due to truncation. Following recent studies on long-context modeling as LongEmbed~\cite{zhu-etal-2024-longembed}, the segmentation parameters are empirically determined to achieve a balance between contextual completeness and computational efficiency. Specifically, we set the window size to $1500$ characters to provide sufficient contextual coverage without exceeding the input capacity of the dense model, and the overlap is set to $300$ characters, corresponding to about $20\%$ of the window size, to maintain semantic continuity. The fragments, which are shorter than $100$ characters, are removed to avoid semantically insignificant segments. 
Each valid segment is then converted into a high-dimensional embedding vector using the semantic model, assigned a unique identifier and metadata, and subsequently batch-uploaded to the vector database.
 
\subsubsection{Dense retrieval strategy}

After generating the query vector, the RAG system retrieves relevant contextual information from the knowledge base through dense retrieval. The query vector is submitted to a vector database specifically designed for high-dimensional similarity searches, where cosine similarity is used to compare the query vector against all stored vectors in the knowledge base. The number of relevant vectors $N_{\mathrm{ret}}$ retrieved in each search depends on the length of the query contract and the number of its segmented blocks.
Specifically, given the length of a query contract $L_q$, a window size $s$, and an overlap $o$, the number of retrieved vectors $N_{\mathrm{ret}}$ can be expressed as
\begin{equation}
N_{\mathrm{ret}} = \chi\left\lceil \frac{L_q - o}{s - o} \right\rceil,
\end{equation}
where $\chi$ denotes the number of top-ranked vectors returned for each segment. The retrieved fragments inherit the vulnerability labels from the original contracts, and the final vulnerability prediction is obtained through a label aggregation mechanism. The dense retrieval score is computed based on the cosine similarity~\cite{zhao2024dense}, which is given by
\begin{equation}
    \mathrm{Score}\left({\boldsymbol{q}},{\boldsymbol{d}} \right) =\frac{{\boldsymbol{q}}\cdot {\boldsymbol{d}}}{\left\| {\boldsymbol{q}} \right\| \cdot \left\| {\boldsymbol{d}} \right\|}.
\end{equation}

To ensure the independence of aggregation, fragments originating from the query contract are excluded from the retrieval results. The vulnerability labels of the remaining fragments are aggregated through a dynamic voting mechanism, where label frequency reflects its relative evidence across retrieved fragments. Unlike fixed-threshold aggregation, we adopt an adaptive thresholding strategy that dynamically adjusts based on the total number of retrieved fragments. This approach effectively reduces noise from semantically similar but non-vulnerable code segments, thereby enhancing the robustness of multi-label vulnerability detection in smart contracts. Specifically, the dynamic voting threshold $\mathbf{\tau}^*$ is defined as the greater of $40\%$ of the retrieved fragments or a minimum threshold of $1$, which is given by
\begin{equation}
    \mathbf{\mathbf{\tau}} ^*=\max \left( N^*\times 0.4,1 \right),
\end{equation}
where $N^*$ represents the total number of retrieval results.

This adaptive approach systematically accounts for variations in contract complexity and length, ensuring that both simple contracts, which generate relatively few fragments, and complex contracts, which produce numerous fragments, are subject to appropriately calibrated thresholding. The votes corresponding to each vulnerability label are then aggregated across all retrieved fragments, and only the labels whose vote counts exceed the dynamic voting threshold are retained in the final prediction. Specifically, during aggregation, each retrieved fragment casts one vote for its associated vulnerability labels. The dynamic threshold $\mathbf{\tau}^*$ specifies the minimum number of supporting votes required for a label to be retained. A higher threshold enforces stronger consensus and reduces false positives, whereas a lower threshold increases sensitivity but may introduce spurious labels. The adaptive design of $\mathbf{\tau}^*$ balances these trade-offs according to the retrieval volume.

\subsection{Result Outputs}

The hybrid RAG system ultimately outputs all vulnerability types detected for the current smart contract. These results are then comprehensively evaluated alongside the outputs of the LLM during the subsequent verification module, thereby enhancing the overall reliability and coverage of vulnerability detection.
When executed in parallel, the two detection pathways complement each other through distinct retrieval mechanisms. The dense retrieval pathway captures deep semantic and contextual vulnerability patterns, such as logical inconsistencies and hidden contract dependencies~\cite{zhao2024dense}, while the BM25-based pathway focuses on lexical matching to identify explicit code-level vulnerabilities~\cite{robertson2025bm25}. This integration provides a unified representation that combines semantic and logical perspectives, thereby enhancing both detection coverage and interpretability of the hybrid RAG system.

\section{Meta-learning gated Verification Module}
\label{section: VI}

In this section, we propose a meta-learning gated verification module to enhance the reliability of vulnerability detection results. This module takes as input the detection outputs from both the LLM and the hybrid RAG system. By leveraging meta-learning techniques~\cite{lin2021metagater}, the module can rapidly adapt to new tasks and effectively learn the importance weights of different features, thereby improving the robustness and generalization of the final verification process.

\subsection{Dynamic Feature Weighting}

The meta-learning model comprises two key components: a meta-learner and a base learner. The meta-learner guides the optimization of the base learner, enabling rapid adaptation to new tasks even with limited labeled data. Through this hierarchical learning structure, the input features are dynamically reweighted based on their learned importance. Specifically, we denote $w$ as the feature-weight vector of the meta-learner and $\tilde{w}_t$ as the task-specific weight vector for detection task $t$~\cite{lin2021metagater}. The adaptation process can be formulated as~\cite{lin2021metagater}
\begin{equation}
    \tilde{w}_t = \argmin_{\tilde{w}}\mathcal{L}_t(\tilde{w}) + \frac{\lambda}{2} \| \tilde{w} - w \|_2^2,
\end{equation}
which constrains the parameters of the base learner to remain close to the initialization of the meta-learner while allowing task-specific adjustments. The regularization coefficient $\lambda$ controls the balance between stability and adaptability, namely a larger $\lambda$ enforces stronger adherence to the prior knowledge of the meta-learner, whereas a smaller $\lambda$ promotes faster adaptation to new task-specific distributions.
By leveraging knowledge from historical tasks, the meta-learning model automatically adjusts feature weights, enabling it to swiftly identify and emphasize the most relevant features when encountering new tasks.

\subsection{Model Training}

The meta-learning model is capable of extracting representative feature weights by learning from multiple related tasks~\cite{mao2022metaweighting}. These learned weights are then applied to process features in the current task, thereby enhancing detection accuracy. Furthermore, the model adaptively adjusts these weights according to the feature distributions of different tasks~\cite{iwata2025meta}. This dynamic feature-weight adjustment mechanism enhances the adaptability of the meta-learning model across diverse detection scenarios, effectively reducing the risks of false positives and negatives, and ultimately enhancing the overall performance of smart contract vulnerability detection. 

To train the model, we first define the training dataset as
\begin{equation}
    \mathcal{D} = \{(\tilde{\boldsymbol{x}}_i, y_i)\}_{i=1}^{N},
\end{equation}
where $\tilde{\boldsymbol{x}}_i \in \mathbb{R}^d$ denotes the $d$-dimensional original feature vector of the $i$-th sample. The goal of the meta-learning model is to learn a mapping, which is given by
\begin{equation}
    \mathcal{F}_{\boldsymbol{\theta}}: \mathbb{R}^d \rightarrow [0,1],
\end{equation}
where $\boldsymbol{\theta}$ is the parameters of the meta-learner. This mapping predicts the probability of a vulnerability for each input sample, while enabling rapid adaptation to new tasks with limited labeled data.

\subsection{Feature Construction}

For each sample, the prediction results from $\Psi$ base models are aggregated into
\begin{equation}
    \hat{\mathbf{y}}_i = \left[\hat{y}_i^{(1)}, \hat{y}_i^{(2)}, \ldots, \hat{y}_i^{(\Psi)} \right]^\top \in \mathbb{R}^\Psi.
\end{equation}
In this study, we set $\Psi = 3$, where the three base models correspond to the dense retrieval model, the BM25 model, and the SLoRA model, respectively. The input to the meta-learner is subsequently formed as a weighted combination of the prediction outputs from these base models, which is expressed as~\cite{lin2021metagater}
\begin{equation}
    \boldsymbol{x}_i = \mathbf{w} \odot \hat{\mathbf{y}}_i,
\end{equation}
where $\mathbf{w} = [w_1, w_2, w_3]^\top$ denotes the learnable weight vector, and $\odot$ represents element-wise multiplication, which quantifies the relative contribution of each base model to the final meta-feature representation.

\subsection{Meta-Learner Architecture}

The meta-learner is implemented as an MLP consisting of two hidden layers and the final layer:
\begin{align}
&\mathbf{h}^{(1)} = \sigma \left( \mathbf{W}^{(1)} \boldsymbol{x}_i + \mathbf{b}^{(1)} \right), \\
&\mathbf{h}^{(2)} = \sigma \left( \mathbf{W}^{(2)} \mathbf{h}^{(1)} + \mathbf{b}^{(2)} \right), \\
&\hat{y}_i = \mathrm{sigmoid} \left( \mathbf{W}^{(3)} \mathbf{h}^{(2)} + \mathbf{b}^{(3)} \right),
\end{align}
where $\sigma(\cdot)$ denotes the ReLU activation function, $\hat{y}_i \in [0,1]$ represents the predicted probability, and $\mathbf{W}^{(l)}$ and $\mathbf{b}^{(l)}$ denote the learnable weights and biases of the $l$-th layer, respectively. This architecture facilitates non-linear integration of the base model predictions, enabling the meta-learner to capture complex feature interactions and thereby enhance the overall accuracy of vulnerability detection.

\subsection{Objective Function}

Similar to the training objective of SLoRA, we optimize the meta-learner by minimizing the BCE loss, defined as
\begin{equation}
    \mathcal{L} = - \frac{1}{N} \sum_{i=1}^{N} \Big[ y_i \log \hat{y}_i + (1 - y_i) \log (1 - \hat{y}_i) \Big].
\end{equation}
The learnable parameters $\boldsymbol{\theta} = \{ \mathbf{W}^{(l)}, \mathbf{b}^{(l)}\}$ are optimized as
\begin{equation}
   \boldsymbol{\theta}^* = \argmin_{\boldsymbol{\theta}}\mathcal{L}(\boldsymbol{\theta}).
\end{equation}

Based on the above formulation, we design a meta-learning model that integrates the prediction results of multiple base models through a trainable meta-learner, enabling adaptive fusion of features and improved classification accuracy.

\subsection{Computational Complexity Analysis}

The computational complexity of the proposed meta-learning model mainly arises from the forward and backward propagation of the meta-learner during training, and the inference process that aggregates the outputs of the base models.

\subsubsection{Training computational complexity}
Considering that each input feature vector $\boldsymbol{x}_i \in \mathbb{R}^{\Psi}$ is processed by a two-layer MLP with hidden dimensions $h_1$ and $h_2$, the forward propagation of the meta-learner requires $\mathcal{O}\!\left(\Psi h_1 + h_1 h_2 + h_2\right)$ operations per sample, and the backward propagation introduces the same order of computation. 
Therefore, for $N$ samples in the training dataset, the total training computational complexity is $\mathcal{O}\!\left(N (\Psi h_1 + h_1 h_2 + h_2)\right)$, which scales linearly with both the dataset size $N$ and the number of base models $\Psi$. Since both $\Psi$ and the hidden layer dimensions are small, the additional training overhead of the meta-learner is negligible relative to that of the base models.

\subsubsection{Inference computational complexity}
During inference, each base model independently produces its prediction, 
with a total computational cost of $\sum_{\phi=1}^{\Psi} \mathcal{O}(\mathcal{C}_\phi)$, 
where $\mathcal{C}_\phi$ denotes the inference complexity of the $\phi$-th base model. 
The meta-learner then performs a lightweight fusion operation, requiring $\mathcal{O}\!\left(\Psi h_1 + h_1 h_2 + h_2\right)$ operations per prediction. Hence, the overall inference complexity can be expressed as
\begin{equation}
    \mathcal{O}\!\left(\sum_{\phi=1}^{\Psi} \mathcal{C}_\phi + \Psi h_1 + h_1 h_2 + h_2\right).
\end{equation}
Given the compact structure of the MLP, the additional cost of meta-level fusion remains minimal, 
allowing efficient real-time integration of base model outputs in smart contract vulnerability detection.

\section{Simulation Results}
\label{section: VII}

In this section, we first provide the experimental setup and security analysis of ParaVul. We then evaluate the effectiveness of SLoRA in enhancing LLM-based smart contract vulnerability detection, followed by a detailed evaluation of the hybrid RAG system under both single-label and multi-label scenarios. Finally, we validate the superiority of the proposed verification module, demonstrating its robustness and reliability in achieving accurate vulnerability detection.

\subsection{Experimental Settings}

We implement ParaVul on a server equipped with an Intel Xeon(R) Gold 6133 CPU and an NVIDIA RTX A6000 GPU, using Python 3.10.14 with CUDA 12.1. The main parameter settings of ParaVul are detailed in Table \ref{Key_Parameters}.

To evaluate the performance of ParaVul in detecting both single-vulnerability and multi-vulnerability smart contracts, we utilize the SC\_UEE\footnote{SC\_UEE is available at {\url{https://github.com/1052445594/SC_UEE}}} and ScrawlD~\cite{yashavant2022scrawld} datasets, which encompass various types of vulnerability collected from real-world blockchain platforms and open-source repositories. To ensure seamless integration and enhance processing efficiency, we standardize the SC\_UEE and ScrawlD datasets into a unified JSON format containing the contract code and corresponding vulnerability labels.

\begin{table}[t]
    \renewcommand{\arraystretch}{1.5}
    \caption{Main Parameter Settings of ParaVul}
    \label{Key_Parameters}
    \resizebox{\columnwidth}{!}{
    \begin{tabular}{m{6.0cm}|>{\centering\arraybackslash}m{1.8cm}} 
    \toprule[1.5pt]
    \rowcolor{gray!10}
    \hline
    \multicolumn{1}{c|}{\textbf{Notations}}  & \multicolumn{1}{c}{\textbf{Definition}} \\ \hline
        Learning rate of adapters $\mathbf{\eta}$~\cite{wang2023cost}  & $5\times 10^{-5}$   \\\hline
        Batch size $B$~\cite{wang2023cost}  & $8$ \\\hline
        Number of epoch $T$~\cite{wang2023cost}  & $5$ \\\hline
        LoRA rank $r$~\cite{hu2022lora} & $8$ \\\hline
        Top-$K$ selection of the BM25 model~\cite{robertson2025bm25} &  $7$ \\\hline
        Top-$K$ selection of the dense model~\cite{zhao2024dense} &  $5$ \\\hline
        Voting threshold of the BM25 model~\cite{robertson2025bm25} &  $4$ \\\hline
        Parameter $k_1$ of the BM25 model~\cite{robertson2025bm25} & $1.5$\\\hline
        Parameter $b$ of the BM25 model~\cite{robertson2025bm25} & $0.9$\\\hline
    \bottomrule[1.5pt]
    \end{tabular}
    }
\end{table}

\begin{table*}[t]
\centering
\caption{Performance of LLMs in Smart Contract Vulnerability Detection}
\label{LLM_Result}
\renewcommand{\arraystretch}{1.5} 
\resizebox{\textwidth}{!}{
\begin{tabular}{ccccccccccc}  
    \toprule[1.5pt]
    \hline
\multirow[c]{2}{*}{\textbf{Models}} 
& \multicolumn{5}{c}{\textbf{Single Labeled}} 
& \multicolumn{5}{c}{\textbf{Multi Labeled}} \\ 
\cmidrule(l){2-6} \cmidrule(l){7-11} 
& \cellcolor{yellow!10}{Accuracy$\uparrow$} 
& \cellcolor{yellow!10}{Recall$\uparrow$} 
& \cellcolor{yellow!10}{Precision$\uparrow$} 
& \cellcolor{yellow!10}{F1-score$\uparrow$}  
& \cellcolor{yellow!10}{Time ($\rm{s}$)$\downarrow$} 
& \cellcolor{yellow!10}{Accuracy$\uparrow$} 
& \cellcolor{yellow!10}{Recall$\uparrow$} 
& \cellcolor{yellow!10}{Precision$\uparrow$} 
& \cellcolor{yellow!10}{F1-score$\uparrow$} 
& \cellcolor{yellow!10}{Time ($\rm{s}$)$\downarrow$}\\
\midrule
\cellcolor{gray!6}GPT-4o & 0.0139  &  0.6852 & 0.1518 & 0.2485 & 4.0833 & 0.0058 & 0.5685 & 0.4006 & 0.4700 & 2.9049\\
\cellcolor{gray!6}LLaMA7b & 0.0008  &  0.3873 & 0.0890 &  0.1448 & 0.7912 & 0.0028  &  0.4206 & 0.2440 & 0.3088 & 0.7886\\
\cellcolor{gray!6}LLaMA7b + QLoRA & 0.7731  &  0.8125 &  0.9043 &  0.8559 & \underline{0.4173} & 0.8173  &  0.9199 & 0.9287 &  0.9243 & \textbf{0.4000}\\
\cellcolor{gray!6}LLaMA7b + QALoRA & 0.7129  &  0.7480 &  0.8616 &  0.8008 & 0.4190 & 0.7420  &  0.8846 & 0.9068 &  0.8956 & \underline{0.4108}\\
\rowcolor{blue!10}
\textbf{LLaMA7b + SLoRA} & 0.8055  &  0.8228 &  0.8931 & 0.8565 & \textbf{0.3998} & 0.7913  &  \underline{0.9237} & 0.9262 &  0.9249 & 0.4140\\\hline
\cellcolor{gray!6}LLaMA13b & 0.0010  & 0.5275 &  0.0890 & 0.1972 & 0.4301  & 0.0115  & 0.4287 &  0.2894 & 0.3455 & 0.8393\\
\cellcolor{gray!6}LLaMA13b + QLoRA &\underline{0.8240}  &  0.8346 &  \underline{0.9257} & 0.8778 & 0.7662 & \underline{0.8318}  &  0.9225 & 0.9365 &  0.9295 & 0.7855\\
\cellcolor{gray!6}LLaMA13b + QALoRA & \underline{0.8240}  &  \underline{0.8492} &  0.9224 & \underline{0.8842} & 0.7891 & 0.8202  &  0.9213 & \underline{0.9404} &  \underline{0.9307} & 0.7745\\
\rowcolor{blue!10}
\textbf{LLaMA13b + SLoRA} 
& \textbf{0.8611} 
& \textbf{0.8759} 
& \textbf{0.9300} 
& \textbf{0.9021} 
& 0.7960 
& \textbf{0.8637} 
& \textbf{0.9294} 
& \textbf{0.9500} 
& \textbf{0.9396} 
& 0.7789\\
\hline
\bottomrule[1.5pt]
\end{tabular}
}
\end{table*}

\subsection{Security Analysis}

ParaVul ensures both security and reliability through its decentralized architecture and adaptive learning design.

\subsubsection{Decentralization} 
ParaVul removes dependence on any single trusted authority. Specifically, the LLM detector and the RAG detector operate independently, while the meta-learning gated verification module fuses their outputs in a decentralized manner, thereby eliminating single points of failure and improving system resilience.

\subsubsection{Integrity and traceability} 
All analyzed contract fragments and detection results are securely stored with unique identifiers in a tamper-resistant vector database, ensuring traceable and reproducible vulnerability analysis. 

\subsubsection{Robustness and privacy} 
In ParaVul, the adaptive fusion mechanism mitigates the influence of biased or adversarial detectors, while sensitive contract data are locally processed or encrypted to prevent information leakage.

Overall, ParaVul provides a secure, interpretable, and trustworthy approach for smart contract vulnerability detection.

\subsection{Performance Evaluation of SLoRA}

As shown in Table~\ref{LLM_Result}, we summarize the performance of multiple LLMs with different fine-tuning techniques on both single-label and multi-label vulnerability detection tasks. The evaluation metrics include accuracy, recall, precision, and F1-score, which comprehensively reflect model effectiveness in smart contract vulnerability detection. From Table~\ref{LLM_Result}, we observe that LLaMA13b with SLoRA achieves state-of-the-art performance across all evaluation metrics, significantly outperforming QLoRA~\cite{dettmers2023qlora} and Quantization-Aware LoRA (QALoRA)~\cite{xu2023qa}. Furthermore, LLaMA13b with SLoRA surpasses both pre-trained GPT-4o and LLaMA7b across all metrics. In particular, it achieves an F1-score of $0.9021$, compared with $0.2485$ for GPT-4o in single-label tasks.

In single-label detection, LLaMA13b with SLoRA achieves an accuracy of $0.8611$, a recall of $0.8759$, a precision of $0.9300$, and an F1-score of $0.9021$. This performance exceeds that of LLaMA13b with QLoRA, which attains an F1-score of $0.8778$, by a margin of $2.8\%$, and surpasses LLaMA13b with QALoRA by $1.9\%$. Although these margins may appear numerically modest, they represent substantial improvements in the context of smart contract analysis, where each percentage point of F1-score corresponds to a considerable reduction in undetected or misclassified vulnerabilities. In multi-label detection, SLoRA further demonstrates its effectiveness, achieving an accuracy of $0.8637$, a recall of $0.9294$, a precision of $0.9500$, and an F1-score of $0.9396$, outperforming QLoRA. In addition, SLoRA achieves a faster detection time of $0.7789$ seconds, compared with $0.7855$ seconds for QLoRA. Furthermore, it surpasses QALoRA with a $0.9\%$ higher F1-score and a shorter inference time. The consistent improvement can be attributed to the architecture design of SLoRA, which integrates low-rank adaptation with sparse structure modeling. This design enables more expressive gradient updates and better parameter utilization under limited fine-tuning budgets, effectively enhancing both accuracy and computational efficiency. In summary, the above results demonstrate that SLoRA effectively enhances both the accuracy and efficiency of LLMs in smart contract vulnerability detection.

As shown in Table~\ref{GPU_Comparison}, we compare the GPU memory consumption across different fine-tuning techniques. Compared with the pre-trained LLMs, SLoRA reduces GPU memory usage by more than $41\%$ during fine-tuning. Furthermore, relative to QLoRA and QALoRA~\cite{xu2023qa}, the increase in GPU memory usage with SLoRA is limited to within $1.1\%$. These results demonstrate that SLoRA not only enhances the performance of LLMs in smart contract vulnerability detection but also maintains high computational efficiency and memory economy during fine-tuning.

\begin{table}[t]
    \centering
    \caption{GPU Memory Usage Comparison for Fine-Tuning Techniques}
    \label{GPU_Comparison}
    {\footnotesize
    \setlength{\tabcolsep}{11pt} 
    \renewcommand{\arraystretch}{1.5} 
    \begin{tabular}{c|cc}  
        \toprule[1.5pt]
        \hline
        \multirow{2}{*}{\textbf{Models}} & \multicolumn{2}{c}{\textbf{GPU Memory Usage (MiB)}} \\ 
        \cmidrule(l){2-3}
         & \cellcolor{yellow!10}\textbf{Single Labeled} & \cellcolor{yellow!10}\textbf{Multi Labeled} \\ 
        \midrule
        \cellcolor{gray!6} LLaMA7b & 16464 & 16452 \\
        \cellcolor{gray!6} LLaMA7b + QLoRA & 8910 & 8910 \\
        \cellcolor{gray!6} LLaMA7b + QALoRA & 8976 & 8976 \\
        \rowcolor{blue!10} \textbf{LLaMA7b + SLoRA} & 8992 & 9012 \\\hline
        \cellcolor{gray!6} LLaMA13b & 24966 & 24966 \\
        \cellcolor{gray!6} LLaMA13b + QLoRA & 14548 & 14586 \\
        \cellcolor{gray!6} LLaMA13b + QALoRA & 14476 & 14474 \\
        \rowcolor{blue!10}
        \textbf{LLaMA13b + SLoRA} & 14676  &14674 \\
        \hline
        \bottomrule[1.5pt]
    \end{tabular}}
\end{table}

\subsection{Performance Evaluation of the Hybrid RAG System}

Table \ref{RAG_Result} presents the performance evaluation of the hybrid RAG system on both single-label and multi-label tasks. Specifically, we observe that in single-label tasks, the dense retrieval approach achieves an F1-score of $0.6851$, higher than the $0.6253$ achieved by BM25, and demonstrates greater accuracy, recall, and precision, indicating superior detection capability. However, BM25 records a much shorter processing time of $0.1190$ seconds compared with $6.1852$ seconds for dense retrieval, reflecting higher time efficiency. For multi-label detection, the dense retrieval method again surpasses BM25, achieving an F1-score of $0.8666$ compared with $0.8327$ for BM25, while also maintaining better accuracy, recall, and precision. Nevertheless, BM25 shows an advantage in efficiency, with a processing time of $0.1180$ seconds compared with $6.3855$ seconds for dense retrieval. Overall, dense retrieval provides stronger performance in both single-label and multi-label vulnerability detection, particularly in identifying multiple vulnerability types, while BM25 offers significantly better time efficiency. Combining these two retrieval strategies enables complementary strengths and provides more reliable auxiliary verification for LLM-based detection.

\begin{table*}[t]
    \centering
    \caption{Performance of the Hybrid RAG System in Smart Contract Vulnerability Detection}
    \label{RAG_Result}
    \renewcommand{\arraystretch}{1.5} 
    \resizebox{\textwidth}{!}{
    \begin{tabular}{ccccccccccc}
        \toprule[1.5pt]
        \hline
        \multirow{2}{*}{\textbf{Retrieval Strategies}} & 
        \multicolumn{5}{c}{\textbf{Single Labeled}} & 
        \multicolumn{5}{c}{\textbf{Multi Labeled}} \\
        \cmidrule(l){2-6} \cmidrule(l){7-11}
        & \cellcolor{yellow!10}{Accuracy$\uparrow$} 
        & \cellcolor{yellow!10}{Recall$\uparrow$} 
        & \cellcolor{yellow!10}{Precision$\uparrow$} 
        & \cellcolor{yellow!10}{F1-score$\uparrow$} 
        & \cellcolor{yellow!10}{Time ($\rm{s}$)$\downarrow$} 
        & \cellcolor{yellow!10}{Accuracy$\uparrow$} 
        & \cellcolor{yellow!10}{Recall$\uparrow$} 
        & \cellcolor{yellow!10}{Precision$\uparrow$} 
        & \cellcolor{yellow!10}{F1-score$\uparrow$} 
        & \cellcolor{yellow!10}{Time ($\rm{s}$)$\downarrow$} \\
        \midrule
        \cellcolor{gray!6} Dense & 0.5694 & 0.6898 & 0.6804 & 0.6851 & 6.1852 
        & 0.6406 & 0.9213 & 0.8181 & 0.8666 & 6.3855 \\
        \cellcolor{gray!6} BM25 & 0.5833 & 0.5833 & 0.6738 & 0.6253 & 0.1190 
        & 0.6261 & 0.8304 & 0.8349 & 0.8327 & 0.1180 \\
        \hline
        \bottomrule[1.5pt]
    \end{tabular}
    }
\end{table*}

\begin{figure*}[t]
\centering
\subfigure[Performance of the verification module in single-label smart contract vulnerability detection.]
{
    \begin{minipage}[t]{0.36\linewidth}
	\centering
	\includegraphics[width=0.95\linewidth]{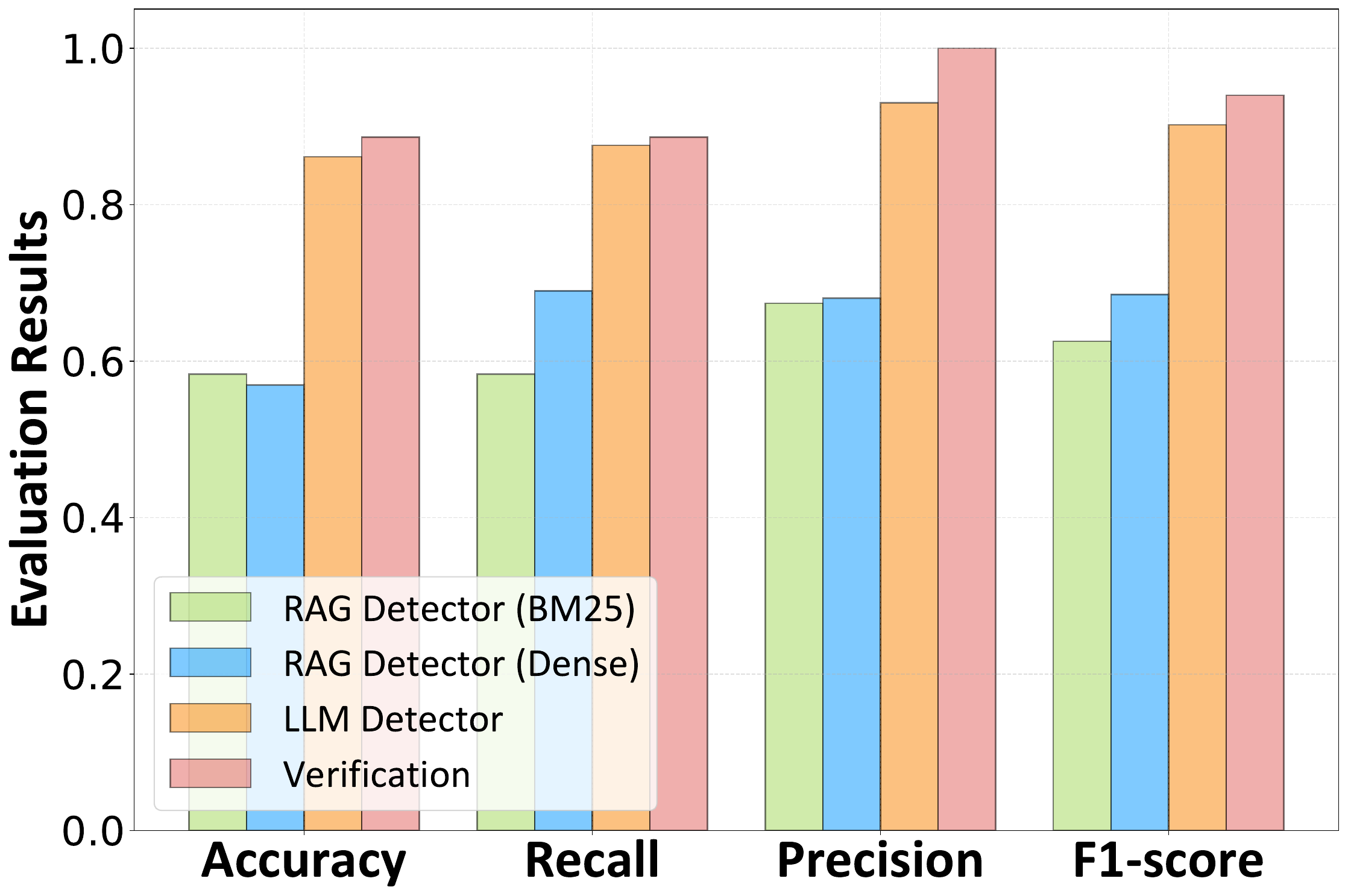}
    \label{single-label_comparision(system)}
    \end{minipage}
}\hspace{0.8in}
\subfigure[Performance of the verification module in multi-label smart contract vulnerability detection.]
{
    \begin{minipage}[t]{0.36\linewidth}
	\centering
	\includegraphics[width=0.95\linewidth]{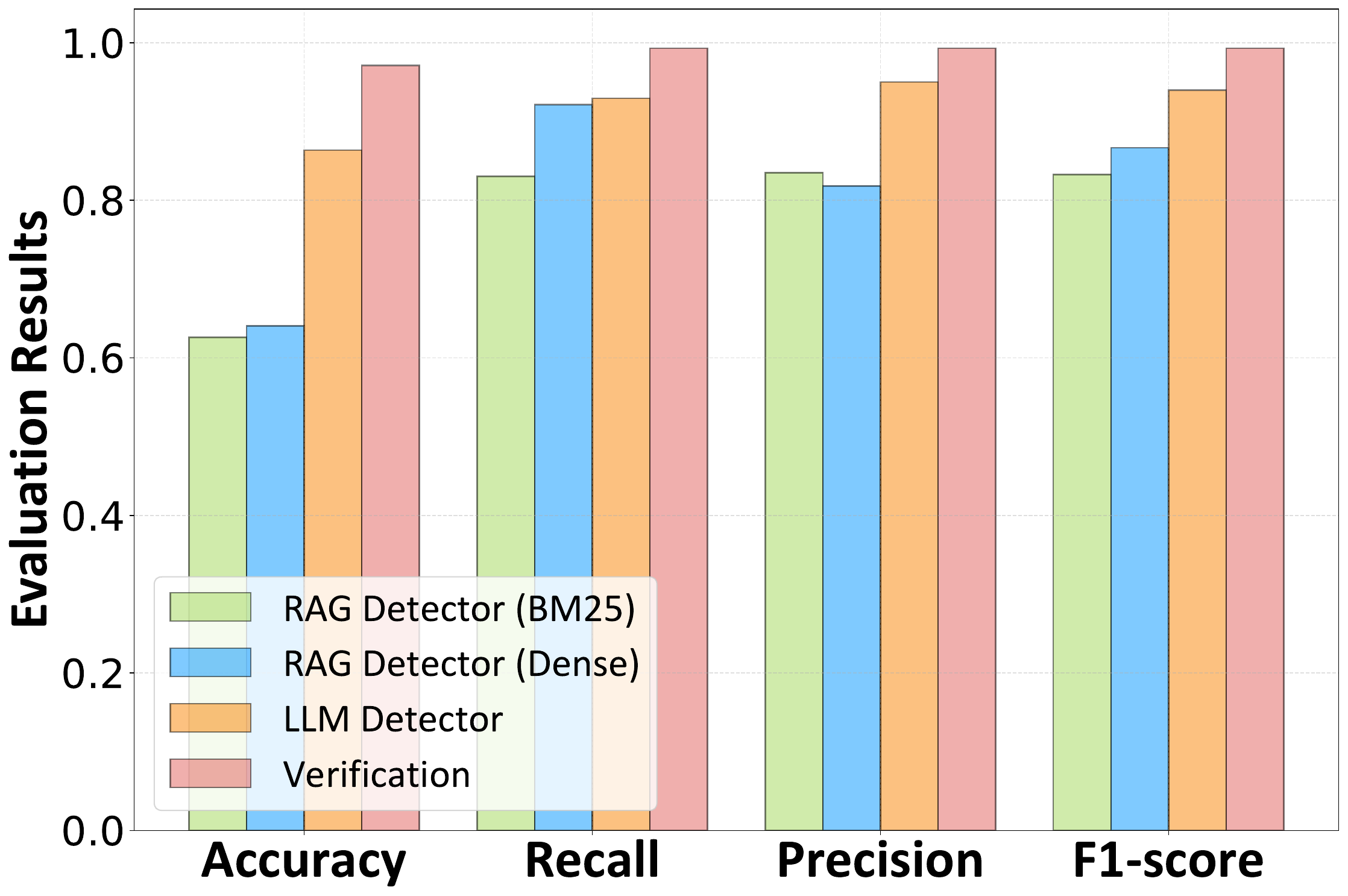}
	\label{multi-label_comparision(system)}
    \end{minipage}
}\hspace{0.8in}
\subfigure[Performance of the meta-learning method in single-label smart contract vulnerability detection.]
{
    \begin{minipage}[t]{0.36\linewidth}
	\centering
	\includegraphics[width=0.95\linewidth]{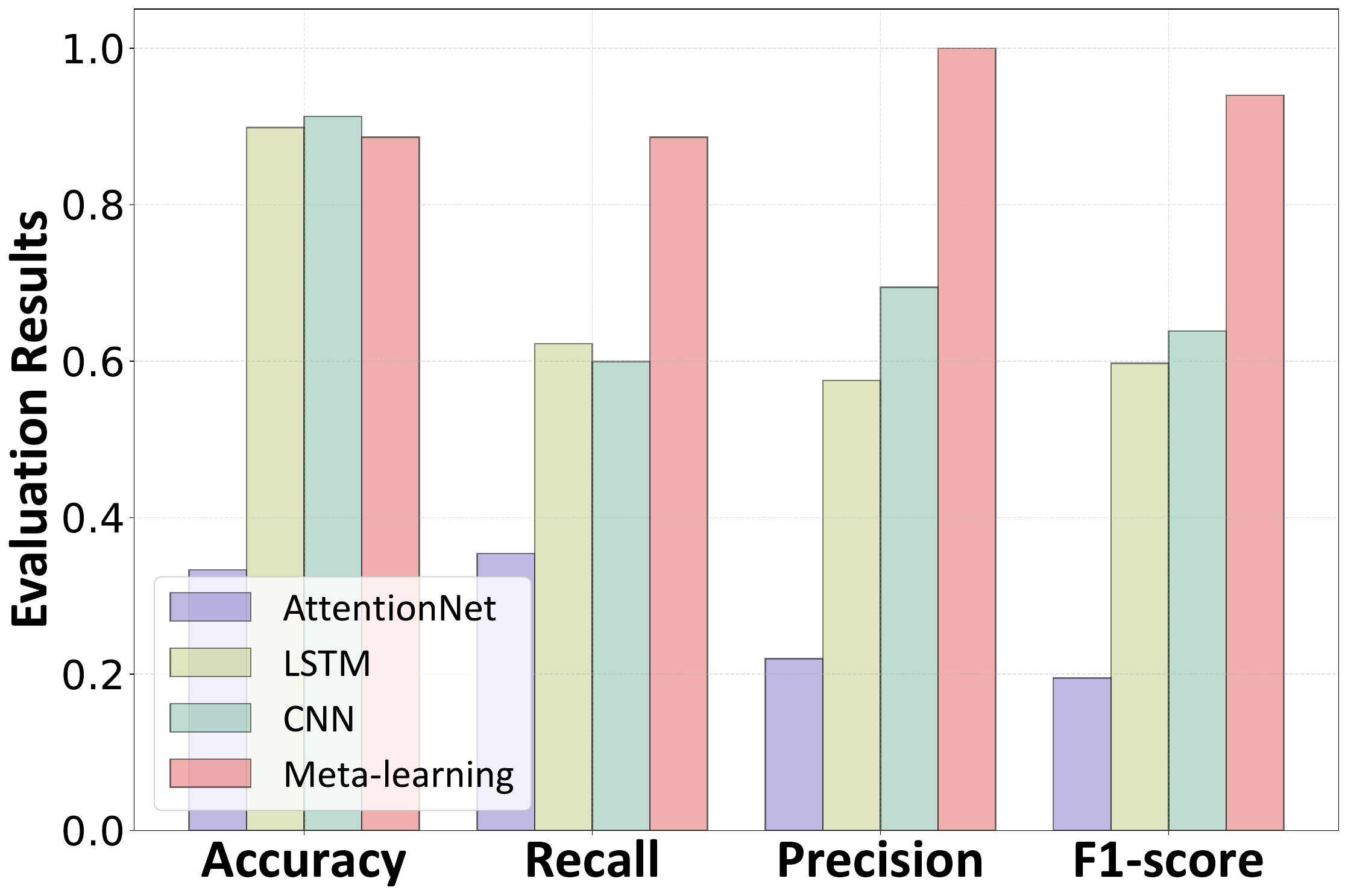}
	\label{single-label_comparision(machine_learning)}
    \end{minipage}
}\hspace{0.8in}
\subfigure[Performance of the meta-learning method in multi-label smart contract vulnerability detection.]
{
    \begin{minipage}[t]{0.36\linewidth}
	\centering
	\includegraphics[width=0.95\linewidth]{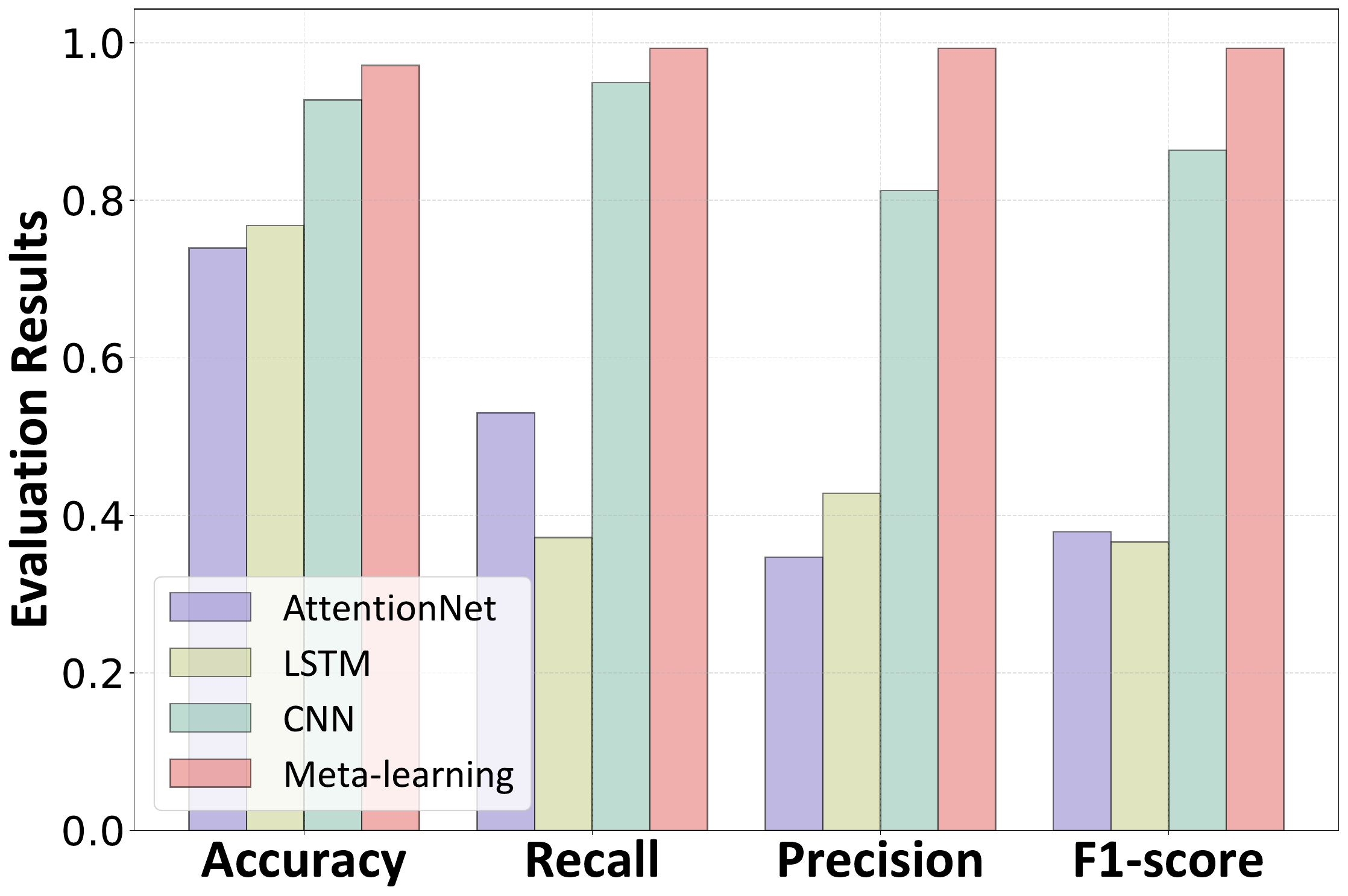}
	\label{multi-label_comparision(machine_learning)}
    \end{minipage}
}
\caption{Performance evaluation of the verification module in smart contract vulnerability detection. Note that the LLM and RAG detectors take raw smart contracts as input, whereas the verification module operates on the intermediate detection outputs generated by these detectors.}
\label{single/muti-label_comparison}
\end{figure*}

\subsection{Performance Evaluation of the Verification Module}

We evaluate the proposed verification module against three baselines: the LLM detector, the RAG detector with BM25 retrieval, and the RAG detector with dense retrieval. As shown in Fig.~\ref{single-label_comparision(system)} and Fig.~\ref{multi-label_comparision(system)}, the module consistently outperforms all baselines in accuracy, precision, recall, and F1-score across both single-label and multi-label tasks. In particular, it achieves substantial gains in precision and recall, demonstrating enhanced robustness and reliability in vulnerability detection. The superior performance of the verification module can be attributed to its meta-learning-based adaptive fusion capability, which effectively integrates the complementary strengths of multiple detectors to achieve more accurate vulnerability predictions. These results confirm that the proposed approach effectively addresses the shortcomings of previous methods and offers a more dependable solution for smart contract security.

As shown in Fig.~\ref{single-label_comparision(machine_learning)} and Fig.~\ref{multi-label_comparision(machine_learning)}, we evaluate the performance of the proposed meta-learning method against traditional machine learning approaches, including AttentionNet~\cite{vaswani2017attention}, Long Short-Term Memory (LSTM), and Convolutional Neural Network (CNN). All baseline models are carefully optimized through grid search on validation data to ensure fair comparison, with hyperparameters such as learning rate, batch size, and hidden dimensions tuned to achieve their best performance. In single-label detection, the meta-learning method achieves an F1-score over $380\%$ higher than AttentionNet, approximately $57\%$ higher than LSTM, and around $47\%$ higher than CNN, with recall also significantly improved. Although CNN attains slightly higher accuracy, the substantial advantages of the meta-learning method in recall and F1-score demonstrate its superior overall effectiveness. In multi-label detection, the meta-learning method achieves an F1-score about $162\%$ higher than AttentionNet, $171\%$ higher than LSTM, and approximately $15\%$ higher than CNN, while maintaining precision and recall near perfect levels. Its accuracy also shows consistent improvement compared with all baseline models. These results confirm the robustness and overall superiority of the meta-learning method, particularly in handling complex multi-label classification tasks.
\begin{figure*}
    \centering
    \includegraphics[width=0.98\textwidth]{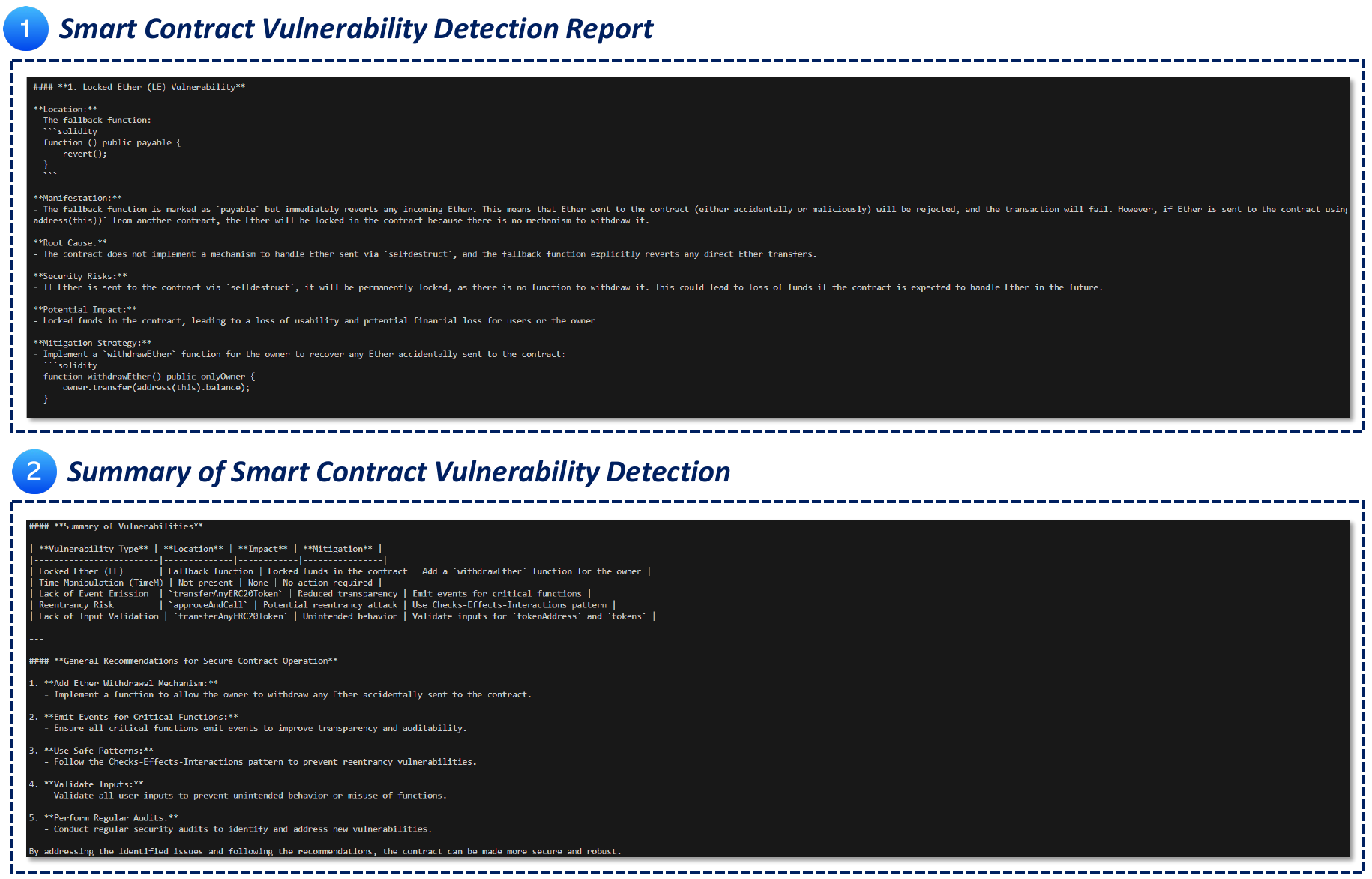}
    \caption{The generated vulnerability detection report comprises two parts: a detailed analysis of each smart contract vulnerability, describing its location, root cause, associated risks, potential impacts, and recommended mitigation strategies; and a structured summary that consolidates all identified vulnerabilities along with their corresponding solution.}
    \label{Report}
\end{figure*}

\subsection{Smart Contract Vulnerability Detection Reports}

As shown in Fig. \ref{Report}, we present a smart contract vulnerability detection report. The generated report provides a comprehensive overview of the detected vulnerabilities. It begins with detailed analyses of individual vulnerabilities, including their location, root cause, associated risks, potential impacts, and recommended mitigation strategies. The findings are then consolidated into a structured summary that highlights vulnerability types, affected code locations, potential consequences, and corresponding mitigation measures. In addition, the report offers general recommendations to enhance the overall security and reliability of the smart contract.

\section{Conclusion}\label{section: VIII}

In this paper, we have proposed ParaVul, a parallel LLM and retrieval-augmented framework for accurate and reliable smart contract vulnerability detection. Specifically, based on QLoRA, we have developed SLoRA to fine-tune LLMs, which effectively improves their detection performance. To enhance the reliability of LLM outputs, we have designed a hybrid RAG system that integrates dense retrieval with BM25. To further improve the detection accuracy of ParaVul, we have introduced a meta-learning gated verification module that performs weighted fusion of LLM and hybrid RAG results. Additionally, we have employed CoT prompt techniques to guide LLMs in generating detailed vulnerability detection reports. Simulation results demonstrate the effectiveness and robustness of ParaVul. For future work, we aim to further optimize SLoRA and develop more advanced verification modules to better handle complex smart contracts, thereby enhancing the overall performance of ParaVul.

\bibliographystyle{IEEEtran}
\bibliography{mybibliography}

\end{document}